\documentclass[10p, conference]{IEEEtran}
\usepackage{cite}
\newcommand{\CLASSINPUTtoptextmargin}{2.4cm}
\usepackage{amsmath,amssymb,amsfonts}
\usepackage{graphicx}
\usepackage{textcomp}
\usepackage{xcolor}
\usepackage{soul}
\usepackage{halloweenmath}
\usepackage{comment}
\usepackage{algorithm}
\usepackage{algpseudocode}
\usepackage{comment}
\usepackage{svg}
\usepackage{multirow}
\def\BibTeX{{\rm B\kern-.05em{\sc i\kern-.025em b}\kern-.08em
    T\kern-.1667em\lower.7ex\hbox{E}\kern-.125emX}}
\algrenewcommand\algorithmicindent{1.0em}
\usepackage{hyperref}
\usepackage{fancyhdr,lipsum}

\fancypagestyle{mahmood}{%
   \fancyhf{} 
   
   \fancyhead[C]{Reprinting or republishing this material for the purpose of advertising or promotion, creating new collective works, reselling or redistributing to servers or lists, or using any copyrighted component in other works must adhere to IEEE policy. The paper has been accepted for publication at \textbf{ICC 2023}, and DOI will be provided as soon as possible.}
}%

\makeatletter
\let\ps@IEEEtitlepagestyle\ps@mahmood
\makeatother

\begin{document}

\makeatletter
\renewcommand{\@biblabel}[1]{[#1]\hfill}
\makeatother

\title{A Scalable Communication Model to Realize Integrated Access and Backhaul (IAB) in 5G}
\author{
\IEEEauthorblockN{Masoud Shokrnezhad}
\IEEEauthorblockA{\textit{CWC, Oulu University, Finland}\\
masoud.shokrnezhad@oulu.fi}
\and
\IEEEauthorblockN{Siavash Khorsandi}
\IEEEauthorblockA{\textit{CEIT, Amirkabir University of Tech., Iran}\\
khorsandi@aut.ac.ir}
\and
\IEEEauthorblockN{Tarik Taleb}
\IEEEauthorblockA{\textit{CWC, Oulu University, Finland}\\
tarik.taleb@oulu.fi}
}

\markboth{This is a pre-printed version.}%
{Shell \MakeLowercase{\textit{et al.}}: Bare Demo of IEEEtran.cls for IEEE Journals}

\maketitle

\begin{abstract}
Our vision of the future world is one wherein everything, anywhere and at any time, can reliably communicate in real time. 5G, the fifth generation of cellular networks, is anticipated to use heterogeneity to deliver ultra-high data rates to a vastly increased number of devices in ultra-dense areas. Improving the backhaul network capacity is one of the most important open challenges for deploying a 5G network. A promising solution is Integrated Access and Backhaul (IAB), which assigns a portion of radio resources to construct a multi-hop wireless backhaul network. Although 3GPP has acknowledged the cost-effectiveness of the IAB-enabled framework and its orchestration has been extensively studied in the literature, its transmission capacity (i.e., the number of base stations it can support) has not been sufficiently investigated. In this paper, we formulate the problem of maximizing transmission capacity and minimizing transmit powers for IAB-enabled multi-hop networks, taking into account relay selection, channel assignment, and power control constraints. Then, the solution space of the problem is analyzed, two optimality bounds are derived, and a heuristic algorithm is proposed to investigate the bounds. The claims are finally supported by numerical results.
\end{abstract}


\IEEEpeerreviewmaketitle

\section{Introduction}\label{s_i}

The strong tides that have shaped digital technologies over the past three decades continue to expand and harden every day. New use cases, such as Unmanned Aerial Vehicle (UAV) based service provision \cite{taleb_supporting_2021, maiouak_dynamic_2019, hossein_motlagh_low-altitude_2016}, holographic communications \cite{makris_cloud_2021}, and extended reality \cite{taleb_toward_2023, taleb_vr-based_2022, taleb_extremely_2021}, have been introduced as the Internet evolves towards a deterministic network of things \cite{qadir_towards_2022, yu_deterministic_2022, yu_deterministic_2022-1, yu_deep_2022, shokrnezhad_near-optimal_2022}. In the near future, due to the fact that billions of devices are anticipated to have stringent quality of service requirements, high-capacity deterministic communication infrastructures must be deployed \cite{nadir_immersive_2021}. 5G, the most recently implemented generation of cellular networks in the telecommunications industry, is one of the potential solutions that constitute a huge technological leap. The heterogeneous architecture employed in 5G enables a large number of access nodes supporting 1000x capacity to operate concurrently within small areas. However, it presents a formidable challenge: increasing the capacity of the backhaul network, which is responsible for connecting radio access components to each other or to the core and transporting signaling messages and data between them. 

\begin{figure}[t!]\centering
\includegraphics[width=3.1in]{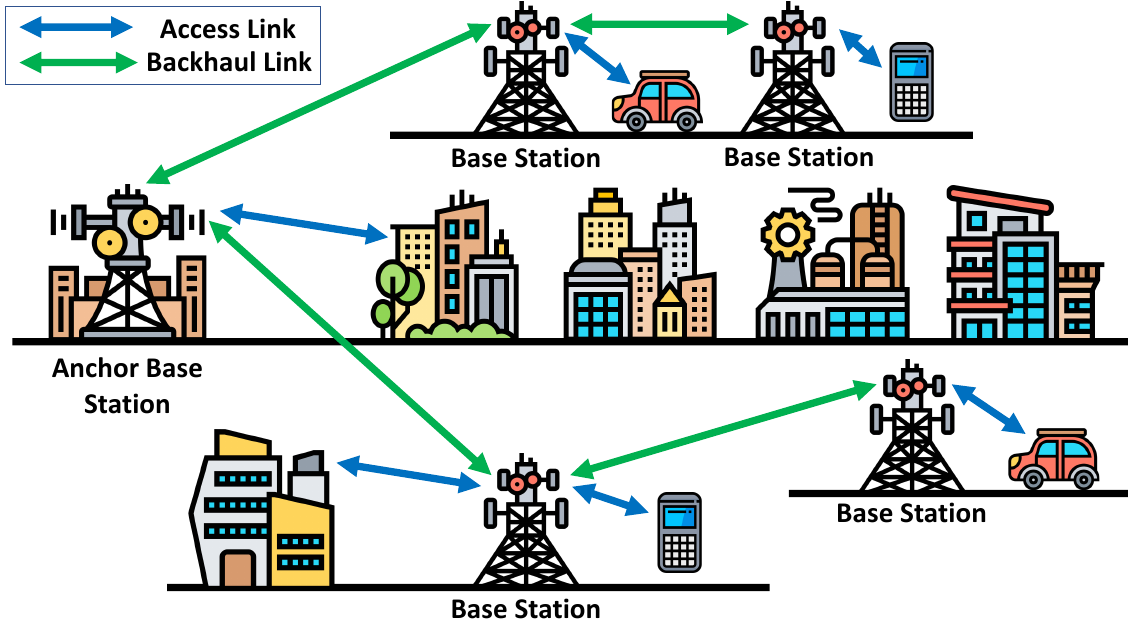}
  \caption{IAB-enabled infrastructure.}
  \label{fig01}
\end{figure}

Integrated Access and Backhaul (IAB) is a futuristic solution wherein only a portion of base stations connect to the infrastructure via fiber, while the others relay the backhaul traffic using wireless links, possibly with multiple hops \cite{zhang_survey_2021}, as illustrated in Fig \ref{fig01}. 3GPP has acknowledged the importance of the IAB-enabled framework as a cost-effective alternative to wired backhaul in a report for 3GPP NR Release 16 \cite{3gpp_nr_2019}, which examines architectures, radio protocols, and physical layer characteristics for sharing radio resources between access and backhaul connections. This study envisions a more advanced and adaptable solution, with support for multi-hop communications, flexible multiplexing of the resources, and a plug-and-play architecture to reduce implementation complexity. Despite widespread agreement that IAB can reduce costs, designing an efficient and high-performance IAB-enabled network remains an open research problem \cite{polese_integrated_2020}.

IAB-enabled networks have been extensively studied in the literature. Liu \textit{et al.} \cite{liu_joint_2020} investigated a resource allocation design in a 5G integrated IAB-enabled network with regard to user fairness. The authors proposed a decomposition-based distributed algorithm and depicted its optimality. Another resource allocation scheme for IAB-enabled networks was presented by Pagin \textit{et al.} \cite{pagin_resource_2022} to increase cell-edge user throughput while decreasing end-to-end delay. Alghafari \textit{et al.} \cite{alghafari_decentralized_2022} proposed a distributed stochastic scheme to jointly solve the problem of bandwidth allocation and path selection in an IAB-enabled multi-hop, multi-path network. Their results showed that the proposed scheme performs almost as well as the optimal centralized algorithm. Lim \textit{et al.} \cite{lim_joint_2021} investigated the joint optimization problem of channel association and power control for multi-hop IAB-enabled networks. They used decomposition techniques and the Lagrangian duality method to solve the problem and demonstrated that configuring multi-hop backhauling improves capacity and coverage. 

Clearly, the majority of previous works aimed to enhance IAB efficiency in terms of various performance metrics, such as fairness, coverage, and bandwidth. However, its transmission capacity has not been adequately investigated. As introduced by Weber \textit{et al.} \cite{weber_transmission_2005} and used in many other research papers \cite{wen_achievable_2013, yang_transmission_2017, lin_stochastic_2014, sheng_transmission_2015}, transmission capacity is the number of base stations that can be supported in terms of their quality of service requirements. This paper fills a gap in the existing literature by formulating the problem of maximizing transmission capacity and minimizing transmit powers for multi-hop IAB-enabled networks taking relay selection, channel assignment, and power control constraints into consideration. The optimality and complexity of the problem are then investigated, and upper and lower limits for transmission capacity and transmit powers are derived. Finally, a heuristic algorithm for solving the problem and investigating its bounds is proposed. 

The remainder of this paper is structured as follows. The system model is explained in Section \ref{s_sm}. In Sections \ref{s_pd} and \ref{s_oa}, the problem definition and optimality analysis are provided, respectively,. Section \ref{s_ras} describes the resource allocation scheme, Section \ref{s_sim} illustrates the results, and Section \ref{s_con} provides concluding remarks.

\section{System Model}\label{s_sm}

Following is a description of the system components examined in this paper: base station placement, channel and propagation models, and quality requirements.

\subsection{Base Station Placement}\label{ss_bsp}

In accordance with the spatial configurations presented by 3GPP \cite{3gpp_nr_2019} for a typical outdoor deployment scenario of a two-tier heterogeneous network, we consider an uplink single-cell cellular network within a bounded two-dimensional region $\mathcal{A}$  and the enclosed area of $\Omega \left( \mathcal{A} \right)$. It is assumed that an Anchored Base Station (ABS) is positioned in the center of the cell and connected to the core network via a high-speed optical fiber. In addition, the network includes $\mathcal{N}$ Small-cell Base Stations (SBSs), whose arrangement is assumed to be the homogeneous Poisson point process with an intensity (or node density) of $\lambda$. The set of SBSs is $\boldsymbol{\mathcal{N}}= \{ 1, \ldots, i, \ldots, \mathcal{N} \}$, $\boldsymbol{\mathcal{M}}$ represents $\boldsymbol{\mathcal{N}} \cup \{\text{ABS}\}$, and $\widehat{\lambda}$ represents the physical limit of the network density. For each $\lambda < \widehat{\lambda}$, it is anticipated that all network characteristics assumed in the remainder of this paper remain viable. In addition, $d_{i,j}$ is the distance between base stations $i$ and $j$ for all $i$ and $j$ in $\boldsymbol{\mathcal{M}}$, and $\boldsymbol{\mathcal{D}}$ is the set of distances. 

\subsection{Channel Model}\label{ss_cham}

To share the spectrum, the Orthogonal Frequency-Division Multiple Access (OFDMA) technique is employed. It is considered that $\mathcal{K}$ isolated resource blocks, dubbed channels and denoted by $\boldsymbol{\mathcal{K}}= \{ 1,2, \ldots, k, \ldots \mathcal{K} \}$, are assigned to backhaul links, and interference between backhaul links (SBS to SBS/ABS) and access links (user device to SBS) is negligible. $\delta _{i}$ indicates the channel of SBS $i$, and the maximum capacity of each link over each channel is $\mathcal{C}$ Mbps.

\subsection{Propagation Model}\label{ss_pm}

The transmit power of SBS $i$ on channel $k$ is denoted by $p_{i,k}$, which is bounded between $0$ and $\widehat{p}$. $ \boldsymbol{p} $ indicates the vector of transmit powers. The received power of SBS $i$ on channel $\delta _{i}$ at its receiver $r_{i}$ (another SBS or ABS) is $\phi _{i,r_i}=p_{i, \delta _{i}}h_{i,r_i}$. In this equation, $ h_{i,r_{i}}= \vartheta d_{i,r_{i}}^{-3} $ represents the path gain from SBS $i$ to the receiver, which is assumed to remain constant during data transmission, where $\vartheta$ is the attenuation factor that represents the power variation due to the shadowing effect \cite{torrieri_principles_2011}, and $d_{i,r_{i}}$ denotes the Euclidean distance between SBS $ i $ and the receiver. $\boldsymbol{\Phi}$ and $\boldsymbol{h}$ indicate the vector of received powers and the matrix of path gains, respectively. For each SBS $i$, there is an interfering sub-network, which is the sphere of radius $\mu_i \left( \lambda \right)$ (the interfering radius), centered at its receiver and denoted by $\varsigma \left(\mu_i \left( \lambda \right) \right)$. $ \phi _{j,r_i} $ of each co-channel SBS $j$ outside of this sphere is less than or equal to $\zeta \sigma^2$, where $\zeta$ is the coefficient controlling the size of the interfering sub-network, and $\sigma ^{2}$ indicates the noise power.

\subsection{Quality Requirements}\label{ss_qr}

SBSs are required to transmit data to ABS at a bit rate of $ \mathcal{R}$ Mbps. To deliver the data flawlessly, a direct connection or a set of multi-hop connections (over other SBSs as relays) should be established between each SBS and ABS. A connection is successful if the sender achieves the minimum required Signal-to-Interference-plus-Noise Ratio (SINR) at the receiver, represented by $\widehat{\gamma}$. The SINR, achieved by SBS $i$ at its receiver $ r_i $ on its assigned channel, is defined as $\gamma _{i,r_{i}}=g \phi _{i,r_{i}} / (\sum _{j \in \boldsymbol{\mathcal{N}} \char`\\ \{i\}, \delta _{j}= \delta _{i}} \phi _{j,r_{i}} + \sigma ^{2})$, where $g$ is the processing gain that is assumed to be identical for all SBSs.

\section{Problem Definition}\label{s_pd}

The main problem is formulated as a Mixed-Integer Non-Linear Programming (MINLP) problem as follows :
\begin{align*}\label{problem}
    &\footnotesize max \sum\nolimits_{\boldsymbol{\mathcal{N}}, \boldsymbol{\mathcal{K}}} \left(\sum\nolimits_{\boldsymbol{\mathcal{M}}} \Lambda_{i,k,m} - \alpha_{i,k}p_{i,k} \right) \tag{\footnotesize OF} \\
    &\footnotesize p_{i,k} \geq \frac{ \Lambda _{i,k,m} \widehat{\gamma}}{h_{i,m}g} \sum\nolimits_{j \in \boldsymbol{\mathcal{N}}, j \neq i} (p_{j,k}h_{j,m}+\sigma^{2}) \quad \begin{aligned}
        &\forall i \in \boldsymbol{\mathcal{N}}, \\
        &\forall k \in \boldsymbol{\mathcal{K}}, \\
        &\forall m \in \boldsymbol{\mathcal{M}}, \\
        \end{aligned}\tag{\footnotesize C1} \\
    &\footnotesize \Lambda _{i,k,m}=r_{i,m,1}x_{i,k} \quad \forall i \in \boldsymbol{\mathcal{N}}, \forall k \in \boldsymbol{\mathcal{K}}, \forall m \in \boldsymbol{\mathcal{M}}, \tag{\footnotesize C2} \\
    &\footnotesize r_{i,m,1} \leq \sum\nolimits_{k \in \boldsymbol{\mathcal{K}}} \Lambda_{i,k,m} \quad \forall i \in \boldsymbol{\mathcal{N}}, \forall m \in \boldsymbol{\mathcal{M}},\tag{\footnotesize C3} \\
    &\footnotesize \sum\nolimits_{k \in \boldsymbol{\mathcal{K}}} x_{i,k} \leq 1 \quad \forall i \in \boldsymbol{\mathcal{N}},\tag{\footnotesize C4} \\
    &\footnotesize \sum\nolimits_{j \in \boldsymbol{\mathcal{N}}} r_{i,ABS,j} \geq 1 \quad \forall i \in \boldsymbol{\mathcal{N}},\tag{\footnotesize C5 } \\
    &\footnotesize \sum\nolimits_{j \in \boldsymbol{\mathcal{N}}} r_{i,m,j} \leq 1 \quad \forall i \in \boldsymbol{\mathcal{N}}, \forall m \in \boldsymbol{\mathcal{M}} \tag{\footnotesize C6} \\
    &\footnotesize \sum\nolimits_{m \in \boldsymbol{\mathcal{M}}} r_{i,m,j} \leq 1 \quad \forall i, j \in \boldsymbol{\mathcal{N}},\tag{\footnotesize C7} \\
    &\footnotesize \sum\nolimits_{m \in \boldsymbol{\mathcal{M}}} r_{i,m,j-1} \geq \sum\nolimits_{m \in \boldsymbol{\mathcal{M}}} r_{i,m,j} \quad \forall i \in \boldsymbol{\mathcal{N}}, \forall j \in \boldsymbol{\mathcal{N}} \char`\\ \{1\},\tag{\footnotesize C8} \\
    &\footnotesize r_{i,m,j} \leq r_{i,z,j-1}r_{z,m,1} \quad \forall i,z \in \boldsymbol{\mathcal{N}}, \forall m \in \boldsymbol{\mathcal{M}}, \forall j \in \boldsymbol{\mathcal{N}} \char`\\ \{1\},\tag{\footnotesize C9}\\
    &\footnotesize \mathcal{R} \sum\nolimits_{i, j \in \boldsymbol{\mathcal{N}}} r_{i,m,j} \leq \mathcal{C} \quad \forall m \in \boldsymbol{\mathcal{N}}.\tag{\footnotesize C10}
\end{align*}

In this problem, $x_{i,k}$ is a binary variable that equals $1$ if channel $k$ is assigned to SBS $i$ and $0$ otherwise. $p_{i,k}$ is a continuous variable that represents the transmit power of SBS $i$ on channel $k$. $r_{i,m,j}$ is a binary variable equal to $1$ if SBS $m$ is chosen as the $j$th relay for SBS $i$. Otherwise, the value will be $0$. If SBS $i$ is directly connected to ABS, $r_{i,ABS,1}$ will be set to $1$,  while $r_{i,m \neq ABS,1}$ and $r_{i,m,z>1}$ for all $m$ will be $0$. $\Lambda _{i,k,m}$ is a binary variable that equals $1$ only if channel $k$ has been assigned to SBS $i$ and SBS $m$ has been designated as its immediate relay. Obviously, $\Lambda _{i,k,m}=x_{i,k} \times r_{i,m,1}$. The primary objective is to maximize transmission capacity while minimizing the sum of transmit powers. The first sum in the objective function represents the number of supported SBSs successfully assigned by a channel and a receiver. The second sum is the transmit power total. $\alpha _{i,k}$ is a small non-negative number so that the major goal is not affected by the sum of transmit powers. $\alpha_{i,k} \leq 1/\widehat{p}$ is a viable option for all $i \in \boldsymbol{\mathcal{N}}$ and $k \in \boldsymbol{\mathcal{K}}$.

The constraints enable the establishment of a single-hop or multi-hop path from each supported SBS to ABS, taking into account the required SINR and transmit power. C1 adjusts the transmit power of each supported SBS based on its assigned channel and next node (the relay to which the supported SBS transmits directly) in order to satisfy its SINR requirement. If SBS $i$ cannot be supported (i.e. $\Lambda_{i,k,m}=0$), the expression on the right-hand side of the inequality will be $0$ and the constraint can be relaxed. C2 defines $\Lambda_{i,k,m}$ as $1$ if and only if $r_{i,m,1}=1$ and $x_{i,k}=1$. C3 guarantees that $r_{i,m,1}$ equals 1 only if SBS $i$ is assigned a channel. This constraint ensures that $r_{i,m,1}$ does not receive unnecessary values if SBS $i$ is not supported (i.e $ \sum_{k \in \boldsymbol{\mathcal{K}}} \Lambda _{i,k,m}=0 $). C4 and C5 guarantee that no more than one channel is assigned to each supported SBS and that at least one path is established from each supported SBS to the ABS, respectively. C6 prevents loops on each SBS's path to ABS. C7 ensures that at each step, each SBS can be assigned no more than one relay. C8 satisfies the condition that the $j$th relay of each SBS is assigned if its $(j-1)$th relay is set. C9 ensures that SBS $i$ selects SBS $m$ as its $j$th relay if another SBS $z$ is selected as $(j-1)$th relay of SBS $i$ and is directly connected to SBS $m$. C8 and C9 guarantee that the established paths are not disjoint. Each SBS's capacity is guaranteed by C10.

\section{Optimality Analysis}\label{s_oa} 

\begin{figure}
\centering
\vspace{0.05in}
\includegraphics[width=3.3in]{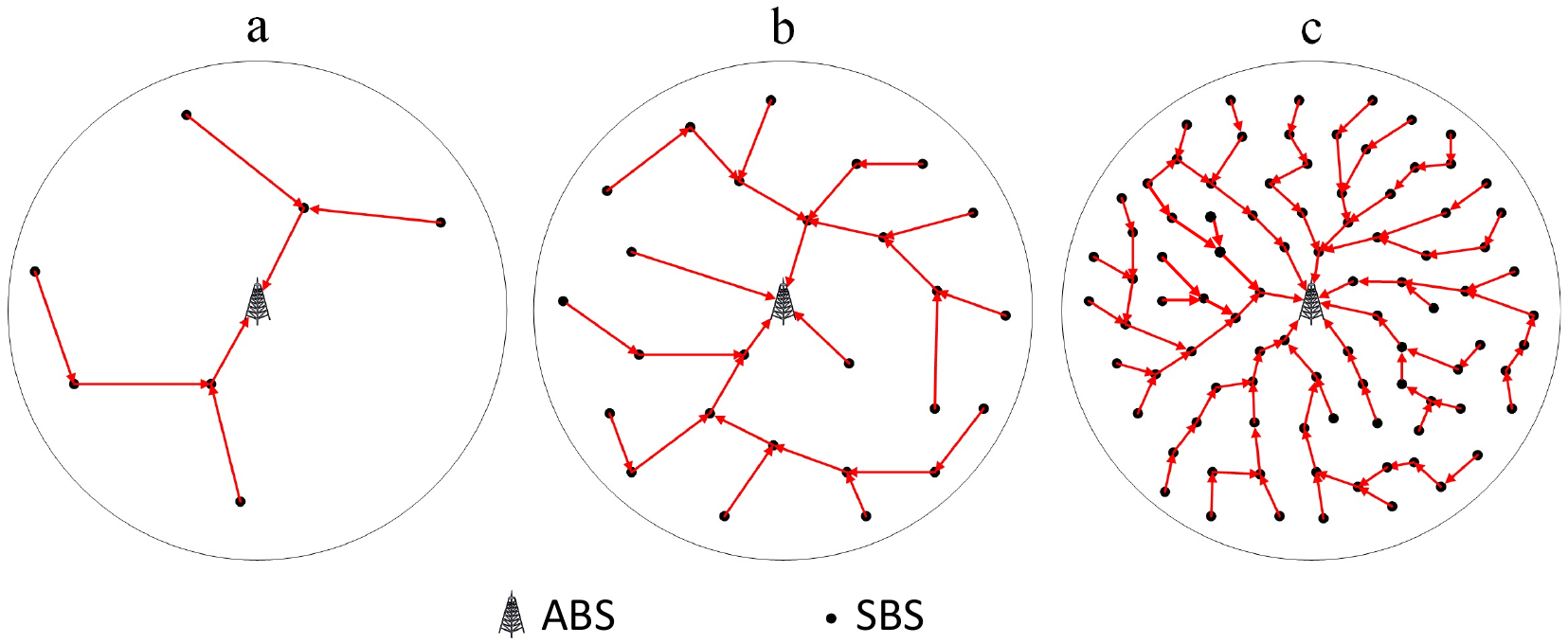}
\caption{MCM for a) $\lambda$, b) $3\lambda$, and c) $9\lambda$.}
\label{fig02}
\end{figure}

This section's aim is to derive optimality bounds for the objective function of the problem defined in Section \ref{s_pd} (i.e., maximizing transmission capacity while minimizing transmit powers). To maximize the objective function, the interference region of transmitters must be confined, which is directly proportional to the connection distance. Therefore, in this section, we first derive connection distances for the optimal communication model, where each SBS is linked to its nearest neighbor SBS (as its relay) so that at least one multi-hop path is established from each SBS to ABS \cite{sun_covertness_2023}, namely Multi-hop Communication Model (MCM). Using the decode-and-forward cooperative model, each relay SBS simultaneously transmits its own data and cooperates in relaying forced by administrative enforcement or incentive mechanisms. Conceptually, the model is depicted in Fig. \ref{fig02}. The number of channels necessary to maximize transmission capacity is then determined based on the calculated distance. Finally, using the determined number of channels, it is proved that MCM is scalable, and optimality bounds for transmit powers and transmission capacity are deduced.

\subsection{Distance to Nearest Neighbor}\label{ss_distant}

As stated previously, the first step is to determine the connection distance between each SBS and its nearest neighbor. Suppose that $\varsigma \left(r \right)$ is the sphere (or disc) of radius $r$ centered at a typical SBS $i$, and $\mathcal{N} \left( \mathcal{A}' \right)$ is the number of SBSs in sub-region $\mathcal{A}'$ for any $\mathcal{A}' \subseteq \mathcal{A}$. According to Moltchanov \textit{et al.} \cite{moltchanov_distance_2012}, \textit{the probability of the specified disc containing $n$ SBSs} is as follows:
\begin{equation}\label{eq_prob1}\footnotesize
\mathcal{P} \left( \mathcal{N} \left( \varsigma \left(r \right) \right) =n \right) = \frac{\left( \lambda \pi r^{2} \right) ^{n}}{n!} e^{- \lambda \pi r^{2}}.
\end{equation}
Given this, \textit{the probability of there being at least $n$ SBSs within disc $\varsigma \left(r \right)$} is:
\footnotesize
\begin{align}\label{eq_prob2}
\begin{aligned}
    \mathcal{P} \left( \mathcal{N} \left( \varsigma \left(r \right) \right) \geq n \right) & = 1- \sum\nolimits_{j=0}^{n-1}\mathcal{P}\left( \mathcal{N} \left( \varsigma \left(r \right)\right) =j \right)\\
    & = 1- \left( e^{- \lambda \pi r^{2}}+ \ldots +\frac{ \left( \lambda \pi r^{2} \right) ^{n-1}}{ \left( n-1 \right) !}e^{- \lambda \pi r^{2}} \right),
\end{aligned}
\end{align}
\normalsize
and when this probability is differentiated with respect to $r$, \textit{the Probability Density Function (PDF) of the distance between the typical SBS and the $n$th nearest SBS} is obtained, that is:
\footnotesize
\begin{align}\label{eq_pdf}
\begin{aligned}
    \mathcal{F}_n \left( r \right) & = \frac{ \partial \mathcal{P} \left( \mathcal{N} \left( \varsigma \left(r \right) \right) \geq n \right) }{ \partial r} \\
    & = \frac{ \partial \left( 1- \left( e^{- \lambda \pi r^{2}} + \ldots +\frac{ \left( \lambda \pi r^{2} \right) ^{n-1}}{ \left( n-1 \right) !}e^{- \lambda \pi r^{2}} \right) \right) }{ \partial r} \\
    & = \frac{2 \left( \lambda \pi \right) ^{n}}{\left( n-1 \right)!} r^{2n-1}e^{- \lambda \pi r^{2}}.
\end{aligned}
\end{align}
\normalsize
Now, if $d_{n}$ represents \textit{the distance between a typical SBS and the $n$th nearest SBS}, the expected value of $d_{n}$ is:
\begin{equation}\label{eq_dev}\footnotesize
E \left[ d_{n} \right] = \int _{0}^{\infty} r\mathcal{F}_{d_{n}} \left( r \right) \textrm{d}r= \int _{0}^{\infty}\frac{2 \left( \lambda \pi \right) ^{n}}{ \left( n-1 \right) !}r^{2n}e^{- \lambda \pi r^{2}}\textrm{d}r,
\end{equation}
and if $ n=1 $, \textit{the distance PDF to the nearest SBS} is simplified to the Rayleigh distribution, $ 2 \lambda \pi re^{- \lambda \pi r^{2}} $, and $ E \left[ d_{1} \right] $ is $1/2~\sqrt[]{ \lambda }$.

\subsection{Optimal Channel Numbers}\label{ss_ocn}

The next step is to compute \textit{the number of channels required to maximize the number of supported SBSs while the transmit powers are minimized}, denoted by $\mathcal{K}^{\star}$. Taking into account the SINR equation and the average distance between neighbors, and given that each SBS communicates with its nearest neighbor as its relay in MCM, \textit{the expected value of the minimum transmit power required to maintain $\widehat{\gamma}$}, denoted by $E \left[ ~ \widehat{p} ~ \right] $, is $\widehat{\gamma} \sigma^2 / (g \vartheta ( 2 ~ \sqrt[2]{ \lambda } )^{3} ).$ Given this, \textit{the expected value of the interfering radius}, indicated by $E \left[ \mu \left( \lambda \right) \right]$, is $( 1/2 ~ \sqrt[2]{ \lambda }) \sqrt[3]{\widehat{\gamma} / \zeta g}$. To maximize the transmission capacity of the network while all SBSs transmit with the expected minimum transmit power, $\mathcal{K}^{\star}$ must equal the number of SBSs within the expected interfering sub-network, that is:
\begin{equation}\label{eq_on}\footnotesize
\mathcal{K}^{\star} = \mathcal{N} \left( \varsigma \left(E \left[ \mu \left( \lambda \right) \right] \right) \right) = \left\lceil \lambda \Omega \left( \varsigma \left( E \left[ \mu \left( \lambda \right) \right] \right) \right) \right\rceil = \left\lceil \frac{ \pi }{4}\sqrt[3]{ \left( \frac{ \widehat{\gamma} }{ \zeta g} \right)^2} ~ \right\rceil
\end{equation}

\subsection{Network Scalability}\label{ss_ns}

Now, considering that the number of available channels is $\mathcal{K}^{\star}$, assume that the network density is increased from $\lambda$ to $\mathcal{M}\lambda$, where $\mathcal{M}$ is a large number. Similar to (\ref{eq_on}), the new network requires $\mathcal{N} \left( \varsigma \left( E \left[ \mu \left( \mathcal{M} \lambda \right) \right] \right) \right)$ channels to satisfy target SINRs by transmitting with the expected minimum powers, where $E \left[ \mu \left( \mathcal{M} \lambda \right) \right]$ is the anticipated interference radius in the new network, that is $(1/2 ~ \sqrt[]{\mathcal{M} \lambda }) \sqrt[3]{ \widehat{\gamma} / \zeta g}$. Consequently, $\mathcal{N} \left( \varsigma \left( E \left[ \mu \left( \mathcal{M} \lambda \right) \right] \right) \right) = \lceil \left( \widehat{\gamma} / \zeta g \right)^{2/3} \pi /4 \rceil$, which is equal to (\ref{eq_on}). Therefore, even in a $\mathcal{M}$ times denser network, the SINR of all SBSs can still be maintained with the same number of channels. This means that, if the number of available channels is large enough, the transmission capacity of MCM is constrained by the bound of $\lambda$, that is $\widehat{\lambda}$.

\subsection{Capacity Upper Bound}\label{ss_cub}

Even though $\widehat{\lambda}$ is a transmission capacity limit, it should be updated in light of the capacity bottleneck of ABS. Since the maximum number of SBSs directly transmitting data to ABS in an OFDMA network is limited by the number of available channels, at most $\mathcal{K}$ paths can be simultaneously established to ABS, and $\lfloor \mathcal{C}/\mathcal{R} \rfloor$ SBSs can send data to ABS through each path. Taking into account only the data rate demand, the maximum number of supported SBSs is $\mathcal{K} \lfloor \mathcal{C}/\mathcal{R} \rfloor$. After determining two possible bounds, it is evident that the transmission capacity of MCM is constrained by the lower one, that is $\min \left\{ \widehat{\lambda}, \mathcal{K} \lfloor \mathcal{C}/\mathcal{R} \rfloor \right\}$. In other words, the capacity is principally determined by three variables: the maximum achievable network density, the maximum channel rate, and the demand of each SBS.

\subsection{Transmit Powers Lower Bound}\label{ss_tplb}

According to Sub-section \ref{ss_ocn}, if the number of channels equals (\ref{eq_on}) in MCM, the expected minimum transmit power required to maintain $\widehat{\gamma}$ is $\widehat{\gamma} \sigma^2 ( 1/ (g \vartheta 2 ~ \sqrt[]{ \lambda } )^{3})$. Given this, for a network with density $\lambda$, the sum of transmit powers will conveniently equal $\Omega \left( \mathcal{A} \right) \widehat{\gamma} \sigma^2 / 8g \vartheta ~ \sqrt[]{ \lambda }$. Consequently, the sum of expected transmit powers is on the order of $\mathcal{O}(1/\sqrt[]{\lambda})$, indicating that it is a decreasing function of network density.

\section{Resource Allocation Scheme}\label{s_ras}

The problem defined in Section \ref{s_pd} is NP-Hard. It is straightforward to demonstrate by reducing the Maximum Induced Subgraph problem to this problem \cite{andersin_gradual_1996}. Given that the size of the solution space for each SBS is $\mathcal{N}(\mathcal{N}+1)\mathcal{K}$ considering its integer variables, the overall size of the problem is on the order of $\mathcal{O}(\mathcal{N}^{\mathcal{N}(\mathcal{N}+1)\mathcal{K}})$. Note that for each SBS $i$, $\mathcal{N}(\mathcal{N}+1)$ and $\mathcal{K}$ are the sizes of $r_{i,m,j}$ and $x_{i,k}$, respectively. When these variables are known, $\Lambda_{i,k,m}$ can be directly assigned, and $p_{i,k}$ can be calculated in polynomial time \cite{shokrnezhad_decentralized_2019}. Therefore, the problem requires at least exponential time to be solved to optimality. So to investigate and validate the bounds provided in Section \ref{s_oa}, we propose a rel\textbf{A}y \textbf{S}ele\textbf{C}tion, channel assignment, and pow\textbf{E}r co\textbf{NT}rol algorithm elaborated in Algorithm \ref{alg}, namely ASCENT.

\begin{algorithm}[b!]
\caption{ASCENT}\label{alg}
\begin{algorithmic}[1]
\State $\boldsymbol{\mathcal{M'}} \gets \{ \text{ABS} \}$, $\boldsymbol{\mathcal{N''}} \gets \{\}$ 
\State $\boldsymbol{\mathcal{N'}} \gets$ sort $\boldsymbol{\mathcal{N}}$ by distance from ABS in ascending order
\State \textbf{for} each base station $i \in \boldsymbol{\mathcal{N'}}$ \textbf{do}
\State \textbar \textbf{} $(i,m) \gets argmin_{m \in \boldsymbol{\mathcal{M'}}} d_{i,m}$
\State \textbar \textbf{ if} the capacity of $m$ is sufficient \textbf{then}
\State \textbar \textbf{} \textbar \textbf{} $r_{i} \gets m$
\State \textbar \textbf{} \textbar \textbf{} update the capacity of base station $m$
\State \textbar \textbf{} $\boldsymbol{\mathcal{N'}} \gets \boldsymbol{\mathcal{N'}}$\textbackslash$\{i\}$, $\boldsymbol{\mathcal{M'}} \gets \boldsymbol{\mathcal{M'}} \cup \{i\}$
\State \textbf{for} each base station $m \in \boldsymbol{\mathcal{M'}}$ \textbf{do}
\State \textbar \textbf{} $\boldsymbol{\rho} \gets \{ i | r_{i} = m \}$, $\delta_i \gets 1 ~ \forall i \in \boldsymbol{\rho}$, $\boldsymbol{\rho'} \gets \{\}$, $t \gets 0$
\State \textbar \textbf{} \textbf{while} $t \leq |\boldsymbol{\rho}|$ \textbf{do}
\State \textbar \textbf{} \textbar \textbf{ if} $t > 0$ \textbf{then}
\State \textbar \textbf{} \textbar \textbf{} \textbar \textbf{} $k' \gets argmin_{k \in \boldsymbol{\mathcal{K}}} \sum\nolimits_{i \in \boldsymbol{\mathcal{N}}, \delta_i = k} \phi_{i,m}$ 
\State \textbar \textbf{} \textbar \textbf{} \textbar \textbf{} $i' \gets argmax_{i \in  \boldsymbol{\mathcal{\rho}} \text{\textbackslash} \boldsymbol{\mathcal{\rho'}}} p_{i,\delta_i}$ 
\State \textbar \textbf{} \textbar \textbf{} \textbar \textbf{} $\delta_{i'} \gets k' $ 
\State \textbar \textbf{} \textbar \textbf{} $t' \gets 0$, $p_{i, \delta_i} \gets \widehat{p} ~ \forall i \in \boldsymbol{\rho} \cup \boldsymbol{\mathcal{N''}}$ 
\State \textbar \textbf{} \textbar \textbf{} \textbf{while} $t' \leq \mathcal{T}$ \textbf{do}
\State \textbar \textbf{} \textbar \textbf{} \textbar \textbf{} $p_{i,\delta_i} \gets min\{ \widehat{p}, \widehat{\gamma}p_{i, \delta_i}/ \gamma_{i, r_i}\} ~ \forall i \in \boldsymbol{\rho} \cup \boldsymbol{\mathcal{N''}}$  
\State \textbar \textbf{} \textbar \textbf{} \textbar \textbf{} $t' \gets t' + 1$  
\State \textbar \textbf{} \textbar \textbf{ if} $t > 0$ \textbf{then}
\State \textbar \textbf{} \textbar \textbf{} \textbar \textbf{} \textbf{if} the sum of transmit powers is increased \textbf{then}
\State \textbar \textbf{} \textbar \textbf{} \textbar \textbf{} \textbar \textbf{} $\delta_{i'} \gets 1 $ 
\State \textbar \textbf{} \textbar \textbf{} \textbar \textbf{} $\boldsymbol{\rho'} \gets \boldsymbol{\rho'} \cup \{i'\}$
\State \textbar \textbf{} \textbar \textbf{} $t \gets t + 1$
\State \textbar \textbf{} $\boldsymbol{\mathcal{N''}} \gets \boldsymbol{\mathcal{N''}} \cup \boldsymbol{\rho}$
\State Calculate $\sum\nolimits_{\boldsymbol{\mathcal{N}}, \boldsymbol{\mathcal{K}}, \boldsymbol{\mathcal{M}}} \Lambda_{i,k,m}$
\State Calculate $\sum\nolimits_{\boldsymbol{\mathcal{N}}, \boldsymbol{\mathcal{K}}} p_{i,k}$
\end{algorithmic}
\end{algorithm}

The algorithm is initialized in its first and second steps. Through steps 3 to 8, relay base stations are assigned and a network is constructed to shorten communication links. This network is constructed iteratively around $\boldsymbol{\mathcal{M'}}$ by selecting the SBS closest to one of the connected base stations in $\boldsymbol{\mathcal{M'}}$ and connecting it to that base station in each iteration if its capacity requirement is met. Once all base stations have been connected, the channel and transmit power of SBSs are allocated through steps 9 to 25. In each iteration, one of the base stations of $\boldsymbol{\mathcal{M'}}$ is fixed, namely $m$, and the set of SBSs that select base station $m$ as their relay is formed, dubbed by $\boldsymbol{\rho}$. Then, for each base station in $\boldsymbol{\rho}$ in descending order of transmit power, the channel with the lowest received power (i.e., interference) at base station $m$ is selected and assigned, and the transmit power of the base stations whose channels are fixed is updated through $\mathcal{T}$ iterations. The procedure for channel assignment and transmit power control is described in depth by Shokrnezhad \textit{et al.} \cite{shokrnezhad_decentralized_2019}. Finally, transmission capacity and the sum of transmit powers are calculated. 

The complexity of the first loop is equal to the complexity of step 4, which is the sum of $|\boldsymbol{\mathcal{M'}}|$ starting from $1$ to $\mathcal{N}$, or $\mathcal{O}(\mathcal{N}^2)$. The complexity of the second loop equals the complexity of steps 18 and 19, that is the sum of $|\boldsymbol{\mathcal{\rho}}|$ times $\mathcal{T}$, or $\mathcal{O}(\mathcal{N})$. Given that the complexity of other steps is constant, it can be inferred that the complexity of the ASCENT algorithm is $\mathcal{O}(\mathcal{N}^{2})$. It is evident that this approach is significantly more efficient than finding the optimal solution to the problem of Section \ref{s_pd}.

\section{Simulations}\label{s_sim}

\begin{figure}
\centering
\vspace{0.05in}
\includegraphics[width=3.3in]{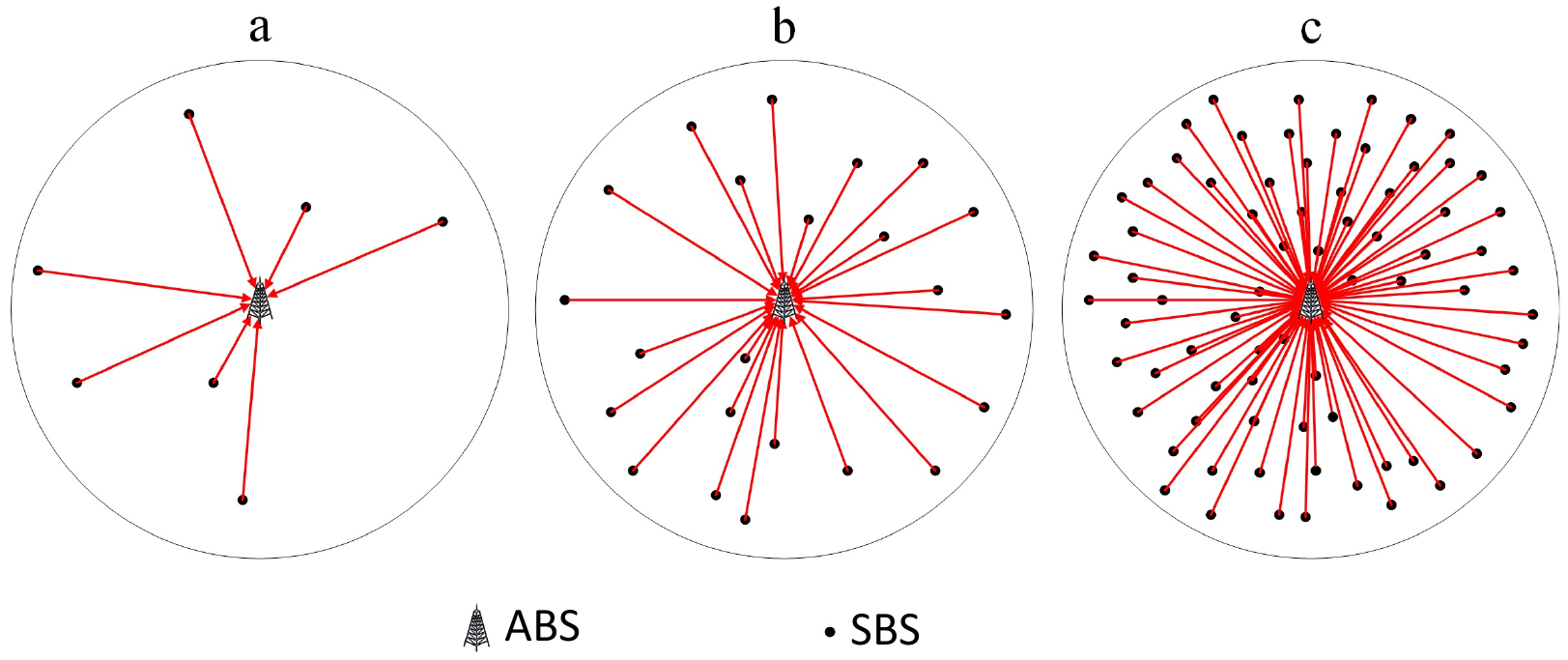}
\caption{SCM for a) $\lambda$, b) $3\lambda$, and c) $9\lambda$.}
\label{fig_scm}
\end{figure}

In this section, the bounds derived for MCM are examined. In order to compare results, the outcomes of the Single-hop Communication Model (SCM) proposed by Shoknezhad \textit{et al.} \cite{shokrnezhad_joint_2018} are also included. In this communication model, SBSs are directly connected to ABS. The model is conceptually illustrated in Fig. \ref{fig_scm} for different densities. The system parameters considered are listed in Table \ref{tab1}. Note that the results were obtained on a computer equipped with an Intel Core i7-4790K processor with a maximum frequency of 4.40 GHz, 8 GB of RAM, and a 64-bit operating system.

\begin{table}[!t]
\centering
\caption{Simulation parameters}
\label{tab1}
\begin{tabular}{c c}
\hline
\pmb{Parameter} & \pmb{Value}\\
\hline
Nodal spatial scattering model & homogeneous Poisson point process \\
Signal propagation model & Omnidirectional \\
ABS cell size & Circular with radius $ 100 m $ \\
SBS cell size & Circular with radius $ 1 m $ \\
$ \widehat{\gamma} $, SINR Requirement & $ 0.3 $ \\
$ \vartheta $, Attenuation factor & $ 0.09 $ \\
$ g $, Antenna gain & $ 1 $ \\
$ \widehat{p} $, Maximum transmit power & $ 2 W $ \\
$ \sigma^2 $, Noise power & $ 10^{-10} W $ \\
\hline
\end{tabular}
\end{table}

\begin{figure}
\centering
\includegraphics[width=\columnwidth]{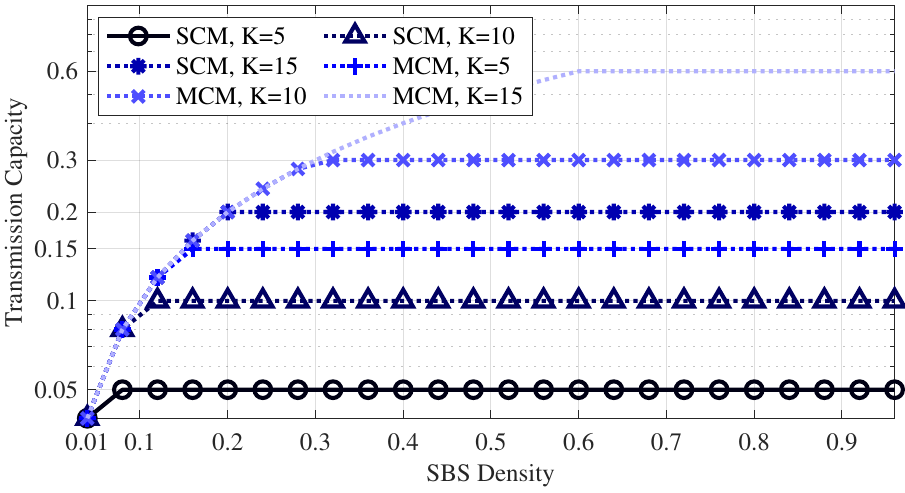}
\caption{Normalized transmission capacity vs. SBS density and the number of available channels (the SINR requirement is $0.3$ and the demand rate of SBSs is $\mathcal{C}/3$).}
\label{fig03}
\end{figure}

Fig. \ref{fig03} depicts the saturation point of transmission capacity versus SBS density and the number of channels for MCM and SCM. The figure shows the normalized transmission capacity, i.e., the number of supported SBSs divided by the size of the cell. As demonstrated, as $\lambda$ increases, the transmission capacity of SCM is limited by the number of available channels ($\mathcal{K}$), which has an inherent upper limit, whereas the upper limit of MCM is $\lfloor \mathcal{C}/\mathcal{R} \rfloor$ times higher. The reason is that increasing network density shortens transmission links and reduces transmit powers and network interference, allowing the multi-hop model to scale effectively. Therefore, Fig. \ref{fig03} substantiates the bounds provided by the mathematical proofs in Section \ref{s_oa}.

\begin{figure}
\centering
\includegraphics[width=\columnwidth]{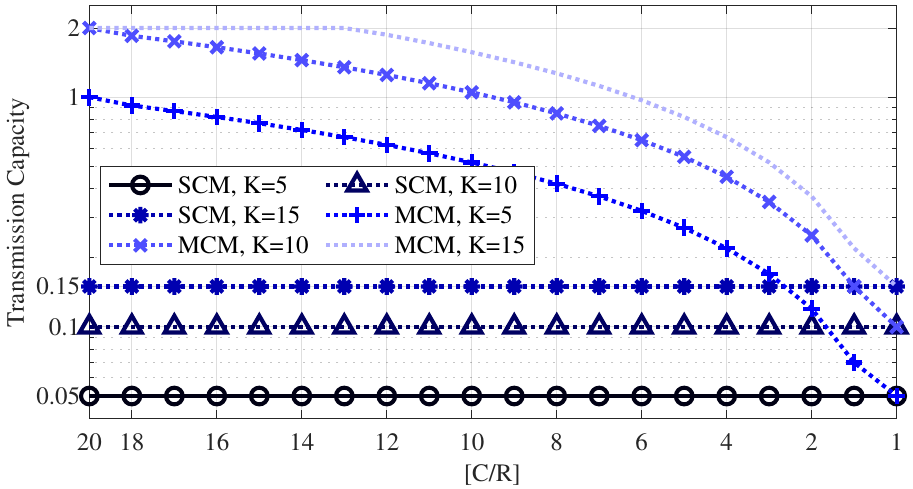}
\caption{Normalized transmission capacity vs. the demand rate of SBSs and the number of available channels (the SINR requirement is $0.3$ and SBS density is $2$).}
\label{fig04}
\end{figure}

Fig. \ref{fig04} compares the normalized transmission capacities for various SBS data rates ($\mathcal{R}$) and a constant channel capacity ($\mathcal{C}$). As demonstrated, increasing $\mathcal{R}$ decreases the MCM transmission capacity, whereas the SCM transmission capacity is independent of the demand rate, and both models converge to the same point when $\mathcal{R}=\mathcal{C}$. It is reasonable, as the number of SBSs whose traffic can be carried by relay SBSs decreases as the demand rate rises. It is evident that the bounds derived in Section \ref{s_oa} are also supported by the results illustrated in this figure.

\begin{figure}
\centering
\includegraphics[width=\columnwidth]{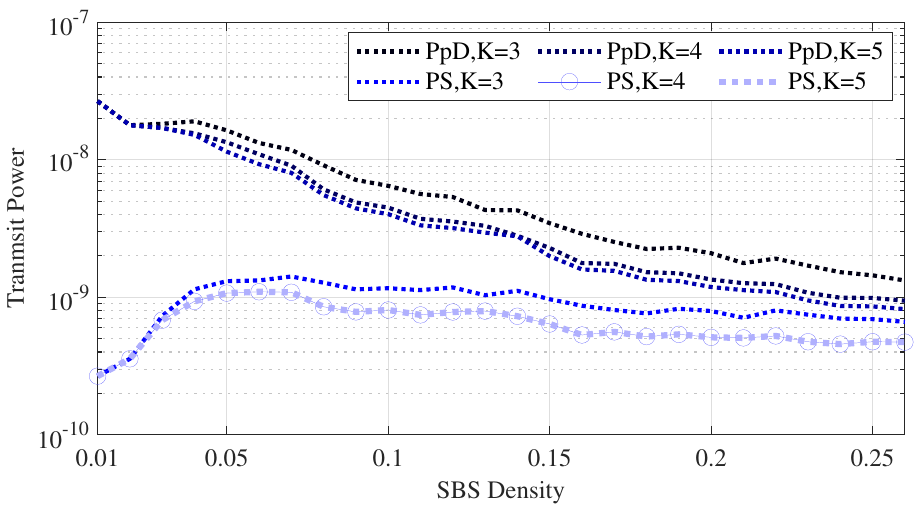}
\caption{Transmit power for MCM vs. SBS density and the number of available channels (the SINR requirement is $0.5$, PpD: Average Transmit Power per SBS Device ($\mathcal{W}$), and PS: Transmit Power Sum ($\mathcal{W}/m^2$)).}
\label{fig05}
\end{figure}

The transmit power consumption in MCM is illustrated in Fig. \ref{fig05} in terms of the average transmit power of each SBS and the total power sum per square meter. As depicted, network densification reduces the transmit power of SBSs, thereby validating the transmit power bound derived in Section \ref{s_oa}. This is due to the fact that increasing the density shortens transmission links, and since there is no interference between immediate neighboring SBSs (using $\mathcal{K}^{\star}$ channels), they require less power to achieve the desired SINR. According to this result, MCM can reduce total energy consumption by adding more SBSs to the network, which can provide network owners with substantial financial and economic benefits. In addition, it is evident from Fig. \ref{fig05} that increasing the number of channels beyond $\mathcal{K}^{\star}$ does not substantially affect the network efficiency.

\begin{figure}
\centering
\includegraphics[width=\columnwidth]{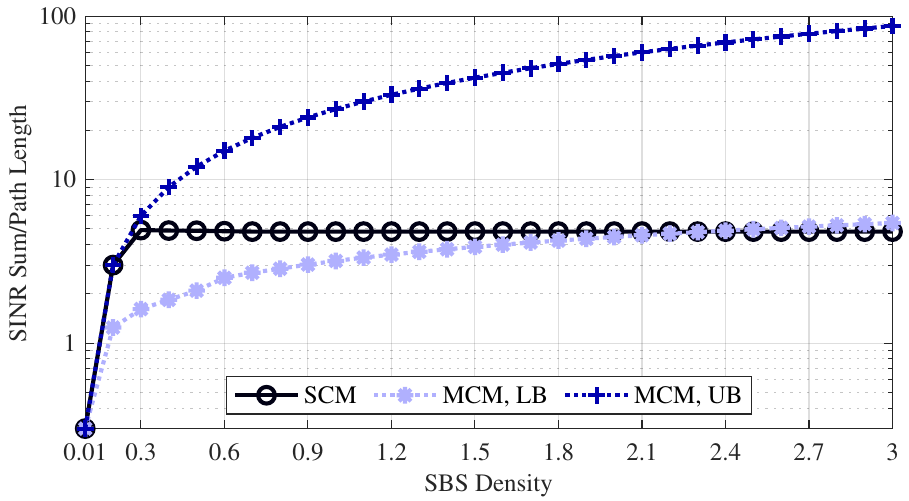}
\caption{The sum of achieved SINRs divided by the path length for SCM and MCM (LB and UB) vs. SBS density (the SINR requirement is $0.3$ and $\mathcal{K} = 3$).}
\label{fig06}
\end{figure}

As the final scenario, the actual network throughput is analyzed while taking into account various communication patterns. In a multi-hop communication setup, the actual network throughput depends on the end-to-end communication pattern. Two extreme instances can be distinguished. In one extreme, all communications utilize single-hop paths (received data blocks in each relay are used to generate new data blocks for transmission to the next receiver), while in the other extreme, all relay SBSs simply transmit data blocks without modification (i.e., the paths are multi-hop originating from SBSs to ABS). These two extreme cases represent, respectively, the Upper Bound (UB) and Lower Bound (LB) of the actual network throughput. As an indication of the actual network throughput in these two extreme cases, Fig. \ref{fig06} depicts the total SINR achieved divided by the average path length. The upper and lower bounds of MCM are increasing, whereas the actual network throughput in SCM remains constant.

\section{Conclusion}\label{s_con}
In this paper, we demonstrated that the backhaul network of 5G can be scaled efficiently in terms of transmission capacity using IAB and applying MCM, in which each small base station communicates with a neighboring station as its relay rather than connecting directly to ABS. First, the problem of maximizing transmission capacity while minimizing transmit powers was formulated while relay selection, channel assignment, and power control constraints were considered. Then, it was demonstrated that the transmission capacity of MCM can be scaled to the physical bound of the base station network density, $\widehat{\lambda}$. In addition, it was demonstrated that in MCM, the sum of transmit powers decreases as network density increases. Finally, a heuristic algorithm for efficiently solving the problem and investigating the derived bounds was proposed, and numerical results supporting the aforementioned claims were presented.

As a potential future direction, the problem of relay selection can be broadened by considering end users' quality of service requirements and network elements' quality of status metrics. For instance, imposing end-to-end reliability and latency constraints drastically reduces the problem's feasible solution space, requiring completely new methods of attack due to the need for solutions with redundant paths (to meet reliability) and shorter lengths (to satisfy latency). Another possible research direction is to replace transmit power with detailed energy consumption functions and cost models in order to customize the backhaul network to accommodate changes in energy providers to minimize energy consumption and energy-related costs, thereby achieving sustainability objectives. A further consideration is the use of machine learning techniques to adapt to the ever-changing nature of future use cases in order to practically implement IAB-enabled networks in beyond-5G systems.

\section*{Acknowledgment}
This work was supported in part by the Academy of Finland 6Genesis project under Grant No. 318927, and the Academy of Finland IDEA-MILL project under Grant No. 352428.

{
\bibliographystyle{IEEEtran}
\bibliography{main}
}

\end{document}